\documentclass[dvips,preprint]{imsart}
\usepackage[dvips]{graphicx}
\usepackage{amsmath,amssymb,amsthm}
\usepackage[latin1]{inputenc}
\usepackage{mathrsfs}
\usepackage{euscript}
\usepackage[english]{babel}
\usepackage{natbib}
\usepackage{a4wide}
\usepackage[usenames,dvips]{color}

\numberwithin{equation}{section}

\begin{document}
\newtheorem{theorem}{Theorem}[section]
\newtheorem{proposition}{Proposition}[section]
\newtheorem{lemma}{Lemma}[section]
\newtheorem{remark}{Remark}[section]

\newcommand\CN{\EuScript{N}}
\newcommand\bmu{\boldsymbol{\mu}}
\newcommand\bSi{\boldsymbol{\Sigma}}
\newcommand\bLa{\boldsymbol{\Lambda}}

\newcommand\bz{\mathbf{z}}
\newcommand\bx{\mathbf{x}}
\newcommand\by{\mathbf{y}}
\newcommand\bA{\mathbf{A}}
\newcommand\bB{\mathbf{B}}
\newcommand\bF{\mathbf{F}}

\begin{frontmatter}

\title{{Testing linear hypotheses in  high-dimensional regressions}}
\runtitle{Testing  in  high-dimensional regressions}



  \thankstext{T3}{The authors acknowledge the support from the
    following research grants: 
NSFC    grant 11171057  (Z. D. Bai),
NSFC    grant 11101181 and RFDP grant 20110061120005 (D. Jiang),
HKU Start-up fund (J. Yao) and 
NSFC  grant NECT-11-0616 (S. Zheng).}
   \begin{aug}
    \author{\fnms{Zhidong} \snm{Bai}\ead[label=e1]{baizd@nenu.edu.cn}}
    \and
    \author{\fnms{Dandan} \snm{Jiang}\ead[label=e2]{jiangdandan@jlu.edu.cn}}
    \and
    \author{\fnms{Jian-feng} \snm{Yao}\ead[label=e3]{{jeffyao@hku.hk}}}
   \and    \author{\fnms{Shurong} \snm{Zheng}\thanksref{T3}\ead[label=e4]{zhengsr@nenu.edu.cn}}

   \runauthor{Z. Bai, D. Jiang, J.  Yao and S. Zheng}

    \affiliation{Northeast  Normal University, Jilin University,
      National University of Singapore,  Universit\'e de Rennes 1, The
      University of Hong Kong}
    \address{Zhidong BAI  and  Shurong ZHENG \\
      KLASMOE and School of Mathematics and Statistics\\
      Northeast Normal University \\
      5268 People's Road\\
      130024 Changchun, China\\
      \printead{e1,e4}
    }
        \address{Dandan JIANG \\
          Institute of Mathematics \\
          Jilin University\\
      2699 Qianjin Street\\
      130021 Changchun, China\\
      \printead{e2}
    }

    \address{Jian-feng YAO \\
     {Department of Statistics and Actuarial Science}\\
      {The University of Hong Kong}\\
      {Pokfulam, Hong Kong}\\
      \printead{e3}
    }
  \end{aug}

\begin{abstract}
For a   multivariate linear model, Wilk's likelihood ratio  test (LRT) constitutes one
of the cornerstone tools. However, the computation of its  quantiles
under the null or the alternative  requires complex analytic
approximations and more importantly, these distributional
approximations are feasible only for moderate dimension of
the dependent variable, say $p\le 20$.
On the other hand, assuming that the data dimension $p$ as well as the
number $q$ of regression variables are fixed while the sample size
$n$ grows, several asymptotic approximations are proposed in the
literature for Wilk's $\bLa$
including the widely used  chi-square approximation.
In this paper, we consider necessary modifications to Wilk's test
in a  high-dimensional context, specifically assuming a high  data
dimension  $p$ and a large sample size  $n$.
Based on recent random matrix theory,
the  correction we propose to Wilk's test
is asymptotically Gaussian under the null
and  simulations demonstrate that the corrected LRT
has very satisfactory
size and power, surely in the large $p$ and large $n$ context,
but also  for moderately large data dimensions like $p=30$ or
$p=50$.  As a byproduct, we give a reason explaining why the standard
chi-square approximation fails for  high-dimensional data.
We also introduce  a   new  procedure for
the classical multiple sample significance test in MANOVA which is valid
for high-dimensional  data.

\vskip1cm
\end{abstract}



\end{frontmatter}

\section{Introduction}

In more and more burgeoning  science and technology fields and with
the help of rapid development in information technology,  a huge
amount of data is collected  where the number of variables  is
usually large. However, most of  traditional statistical tools
deeply depend on the assumption of a large sample size $n$ compared
to the number of variables $p$ (data dimension). For
high-dimensional data analysis, inevitably, these classical tools
become inefficient, or even worse,  inconsistent. For decades,
statisticians devoted special efforts to seek for better approaches
in such high-dimensional data case.  For the   two sample
significance test problem in high dimensions,  as early as in 1958,
\citet{D58} proposed a so-called   {\em non-exact test} (NET) as a
remedy to the failure of  Hotelling's $T^2$-test. A rigorous
analysis of this NET  arises much later in \citet{B96} using
modern random matrix theory (RMT). These authors have found
necessary correction for the $T^2$-test to cope with high
dimensional effects.

Recent work in high dimensional statistics include
\citet{LedoitWolf02}, \citet{Sriv05} and \citet{Schott07}. These
authors propose several procedures in the high-dimensional setting
for testing that i) a covariance matrix is an identity matrix,
proportional to an identity matrix (spherecity) and is a diagonal
matrix or ii) several covariance matrices are equal. These
procedures have the following common feature: their construction
involves some well-chosen distance function between the null and the
alternative hypotheses and rely on the first two spectral  moments,
namely the statistics  tr$S_k$ and  tr$S_k^2$ from sample covariance
matrices $S_k$. In a recent work \cite{B2009}, we have considered
likelihood based tests about  such high dimensional covariance
matrices  where the failure of the classical likelihood ratio test
is explained using RMT. Necessary corrections to these LRT's are
then introduced to achieve consistency.

This paper pursue the investigation of similar questions but for
a multivariate regression  model
with  high dimensional data, i.e.
the dimensions  of the dependent variable as well as the number of the  regression
variables are large compared to the sample size.
More precisely,
let a  $p$-th dimensional regression   model
\begin{equation}\label{linear model}
  \bx_i=\mathbf{Bz}_i+\varepsilon_i,   \quad i=1,\ldots,n
\end{equation}
where $ (\varepsilon_i)$ is a sequence of i.i.d. zero-mean Gaussian
noise $\CN_p(0,\bSi)$ with covariance matrices $\bSi$, $\bB$  a
$p\times q$ matrix of regression coefficients, and $(\bz_i)$ a sequence of known
regression variables of dimension $q$. To simplify the presentation,
we always assume that $n\geq p+q$ and that the rank of $\mathbf{Z}=
(\bz_1, \cdots, \bz_n)$ equals  $q$.

Let us define a {block}  decomposition
$\bB= (\textbf{B}_1, \textbf{B}_2)$ with
$ q_1$ and $q_2$ columns, respectively ($q=q_1+q_2$).
A general linear hypothesis is defined as
\begin{equation}  \label{H02B}      
  H_0 : ~\textbf{B}_1  =  \textbf{B}_1^*~,
\end{equation}
where  $\textbf{B}_1^*$ is a given matrix.
A well-studied example is  the special case $\bB_1^*=0$ yielding
a  significance test for  the first $q_1$ regression variables.

In the general case and under the alternative,
the maximum likelihood estimators of $(\bB, \bSi)$ are
\begin{equation}\label{BMLE}
  \widehat{\textbf{B}}=\left(\sum\limits_{i=1}^n \textbf{x}_i
  \textbf{z}_i'\right)\left(\sum\limits_{i=1}^n \textbf{z}_i
  \textbf{z}_i'\right)^{-1},
\end{equation}
and
\begin{equation}\label{SigmaMLE}
  \widehat{\bSi}=\frac{1}{n}\sum\limits_{i=1}^n
  (\textbf{x}_i-\widehat{\textbf{B}}\textbf{z}_i)(\textbf{x}_i-\widehat{\textbf{B}}\textbf{z}_i)'
.
\end{equation}
The corresponding likelihood maximum equals
\[
  \label{LF}
  \mathscr{L}_1 =(2\pi)^{-\frac{1}{2}pn}|\widehat{\bSi}|^{-\frac{1}{2}n}e^{-\frac{1}{2}pn}.
\]

On the other hand,  under the null hypothesis, by
using a partition
$\textbf{z}'_i=( \textbf{z}'_{i,1},   \textbf{z}'_{i,2})$ on
$q_1$ and $q_2$ variables repectively,
the maximum likelihood estimators  of $(\textbf{B}_2, \bSi)$ are
\begin{equation}\label{B20MLE}
  \widehat{\textbf{B}}_{20}=\left(\sum\limits_{i=1}^n \textbf{y}_i
  \textbf{z}_{i,2}'\right)\left(\sum\limits_{i=1}^n
  \textbf{z}_{i,2} \textbf{z}_{i,2}'\right)^{-1},
\end{equation}
and
\begin{equation}\label{Sigma0MLE}
  \widehat{\bSi}_0=\frac{1}{n}\left(\sum\limits_{i=1}^n
  (\textbf{y}_i-\widehat{\textbf{B}}_{20}\textbf{z}_{i,2})(\textbf{y}_i-\widehat{\textbf{B}}_{20}
  \textbf{z}_{i,2})'
  \right),
\end{equation}
where
$\textbf{y}_i=\textbf{x}_i-\textbf{B}_1^* \textbf{z}_{i,1}$.
The associated  likelihood maximum equals
\begin{equation}   \label{LFw}
  \mathscr{L}_0=(2\pi)^{-\frac{1}{2}pn}|\widehat{\bSi}_0|^{-\frac{1}{2}n}e^{-\frac{1}{2}pn}.
\end{equation}
It follows that the likelihood ratio statistic for the
 test (\ref{H02B}) equals
 \begin{equation}\label{eq:bLa}
   \mathscr{L}_0/  \mathscr{L}_1 = (\bLa_n)^{n/2},
   \quad \bLa_n  =\displaystyle\frac{|\widehat{\bSi}|}{|\widehat{\bSi}_0|}~,
 \end{equation}
where $\bLa_n $ is the celebrated Wilk's $\bLa$ (\cite{Wilks32,
  Wilks34} and \cite{Bart34}).

Let us define a similar  {block} decomposition for the sum
\[   \sum_{i=1}^n \bz_i \bz_i' =
\begin{pmatrix}
  \bA_{11}  &  \bA_{12}\\
  \bA_{21}  &  \bA_{22}
\end{pmatrix},
\]
and the matrix
\[  \textbf{A}_{11:2}=\textbf{A}_{11}-\textbf{A}_{12}\textbf{A}_{22}^{-1}\textbf{A}_{21}.
\]
After  some algebraic manipulations, we get (see \cite{A03}, page 302)
\begin{equation}
  \bLa_n=\left|\textbf{I}+  \frac{q_1}{n-q} \bF \right|^{-1},\label{lamda*}
\end{equation}
where
\begin{equation}\label{Fbf}
  \textbf{F}=\frac{n-q}{q_1}(n\widehat{\bSi})^{-1}(\widehat{\textbf{B}}_1-\textbf{B}_1^*)
  \textbf{A}_{11:2}(\widehat{\textbf{B}}_1-\textbf{B}_1^*)',
\end{equation}
and
$\widehat{\textbf{B}}_1$ is a $p \times q_1$ matrix made of
the first  $q_1$ columns of $\widehat{\textbf{B}}$.

It is known that  $n\widehat{\bSi} \sim W_p(\bSi,n-q)$, a Wishart distribution.
Moreover, under $H_0$,
\[
(\widehat{\textbf{B}}_1-\textbf{B}_1^*)
\textbf{A}_{11:2}(\widehat{\textbf{B}}_1-\textbf{B}_1^*)' \sim
W_p(\bSi,q_1),
\]
and this statistic is independent of $ \widehat{\bSi}$.
Therefore, $H_0$ will be rejected
if $\bLa_n  < \lambda_0$
for some critical value $\lambda_0$, or equivalently, when the matrix $\bF$
has some large enough eigenvalues.

Under the Gaussian assumptions made here,  the exact distribution of
$\bLa_n$ is  known under the null {hypothesis}.
However in practice,   it is usually a difficult task
to compute  the critical value $\lambda_0$
even for moderately large $p$ and $q$.
For example,
\cite{Mathai71} used complex analytical approximations and
established tables for critical values
with $p$ and $q$ smaller than 12.

On the other hand, in a large $n$ asymptotic scheme, one assumes $p$
and $q$ are fixed and then the null distribution of
$-n\log\bLa_n$ is approximated by a $\chi^2_{pq_1}$. Note that
for this chi-squared approximation, one generally uses a rescaled
LRT statistic 
\begin{equation} \label{eq:BBC}
  U_n = -k\log\bLa_n,  \quad k=n-q -\frac12 (p-q_1+1)~.
\end{equation}
This correction is known as Bartlett-Box
correction (hereafter BBC) due to \cite{Box49}
and it is much less biased than the classical  LRT  $-n\log\bLa_n$,
see Section~\ref{sec:simulinear} for a detailed comparison.

However for high dimensional data where
the dimensions $p$ and $q_1$ are large compared to the sample size
$n$, unfortunately the above $\chi^2_{pq_1}$ approximation  becomes
useless.
As an  example, even for moderate  $p$, $q$ and $n$ with
$y_n=p/(n-q)$  close to 1, the  celebrated Mar\v{c}enko-Pastur theorem tell us that
the eigenvalues of $\widehat{\bSi}$
tend to  fill the whole interval $[(1-\sqrt{y_n})^2,~(1+\sqrt{y_n})^2]$.
Hence, a non-negligible proportion of these eigenvalues are close to
zero. Consequently, any statistic based on the inverse
$\widehat{\bSi}^{-1}$ like  $\bLa_n$ becomes unstable and non robust.

In  \S\ref{sec:Clinear}, by using modern RMT,
we introduce a  correction to Wilk's  $\bLa$ to cope with
the mentioned high-dimensional effects.
The corrected LRT  is asymptotically Gaussian and we will see that
it  has very satisfactory
size and power, surely for the large   $p,~q$ and $n$ context,
but also  for moderate  data dimensions like $p=30$ or
$p=50$.

Moreover, to assess the power of the corrected LRT,
we  examine two additional  tests based on statistics of least-squares
type as suggested   in \citet{B96}. A quite intensive simulation
experiment is then conducted to compare these different procedures
for testing  (\ref{H02B}).

Next in \S\ref{sec:Application}, we consider the classical
multiple sample significance test problem but with high-dimensional
data. As it is well-known,  this problem can be embedded into
a special instance
of the general linear hypothesis  (\ref{H02B}).
Therefore, by an application of
general results of  \S\ref{sec:Clinear}, we obtain a valid LRT after
necessary corrections.

All the proofs and technical  derivations are
postponed to   \S\ref{sec:proofs}.

\section{A CLT for linear statistics of random Fisher matrices}
\label{sec:useful}

We first recall a fundamental result from RMT for linear statistics
of so-called random Fisher matrices  which will be used below.
For any $p\times p$ square matrix $M$  with real eigenvalues
$\left(\lambda_i^M\right)$, $F^M$ denotes the  empirical spectral
distribution (ESD) of $M$, that is,
$$
F^M(x) = \frac{1}{p}\sum\limits_{i=1}^{p}\textbf{ 1}_{\lambda_i^M
\leq x}, \quad \quad x \in \mathbb{R}.
$$
We will consider random matrices $(M_n)$ whose ESD $F^{M_n}$
converges, in a sense to be precise and  when $n\to\infty$, to a
limiting spectral distribution (LSD) $F$. Assume we have to estimate
some parameter of $F$, say $\theta = \int f (x)dF(x)$ for some
function $f$, it is natural to use the empirical estimator
$$\widehat{\theta}_n= \int f (x)dF^{M_n}(x)=
\frac{1}{p}\sum\limits_{i=1}^p f(\lambda_i^{M_n}),
$$
which is a  so-called linear spectral statistic (LSS) of the random
matrices $M_n$.

 Let  $ \{\xi_{ki} \in \mathbb{C}, i, k = 1, 2, \cdots
\}$ and $\{\eta_{kj} \in \mathbb{C}, j, k = 1, 2, \cdots \}$ be two
independent double arrays of $i.i.d.$ complex variables with mean 0
and variance 1. Write $\xi_{\cdot i} = (\xi_{1i}, \xi_{2i},\cdots ,
\xi_{pi})^{T}$ and $\eta_{\cdot j} = (\eta_{1j}, \eta_{2j}, \cdots,
\eta_{pj})^{T}$. Also, for any positive integers $n_1, n_2$, the
vectors $ (\xi_{\cdot 1}, \cdots,\xi_{\cdot n_1}) $ and $ (\eta_{\cdot
  1}, \cdots,\eta_{\cdot n_2}) $ can be thought as independent samples of size $ n_1 $
and $ n_2 $, respectively, from some $p$-dimensional distributions.
Let $S_1$ and $S_2$ be the associated sample covariance matrices, ~
 $i. e.$
\[
S_{1}={1\over{n_1}}\sum\limits_{i=1}^{n_1}\xi_{\cdot i}\xi_{\cdot i}^* \quad
\mbox{and}\quad
S_{2}={1\over{n_2}}\sum\limits_{j=1}^{n_2}\eta_{\cdot j}\eta_{\cdot
  j}^*.
\]
Then, the
following  so-called {\em F-matrix}  generalizes  the classical
Fisher-statistic to
the present $p$-dimensional case,
\begin{equation}
V_n=S_{1}S_{2}^{-1}\label{F}
\end{equation}
where we assume that $n_2 > p$. Here we use the notation $n=(n_1, n_2)$.

 Let us also assume that
\begin{equation}
  y_{n_1}=\frac{p}{n_1} \rightarrow y_1 \in (0, 1),  \quad
  y_{n_2}=\frac{p}{n_2} \rightarrow y_2
  \in (0, 1).\label{limitscheme}
\end{equation}
Under suitable moment conditions,  the ESD $F^{V_n}$ of $V_n$
 has a  LSD  $F_{y_1, y_2}$ with the following density function,
 see  p.72 of \cite{B06},
\begin{equation}
  \displaystyle{\ell(x)}=\left\{
  \begin{array}{ll} & \displaystyle{\frac {(1-y_{2})\sqrt{(b-x)(x-a)}}{2\pi x(y_{1}+y_{2}x)},
      \quad  ~a \leq x \leq b,}\\[6mm]
    &\displaystyle{0, \quad\quad \quad\quad \mbox{otherwise},}
  \end{array}
  \right.\label{LSDden}
\end{equation}
where
\[
a=\left( \frac  {1-h}{ 1-y_{2} }\right)^2,\quad
b=\left( \frac  {1+h}{ 1-y_{2} }\right)^2,\quad
h=\sqrt{y_{1}+y_{2}-y_{1}y_{2}}.
\]
Let  $\mathcal{U}$ be an open subset of the complex plane  which
contains   the interval $[a,b]$
and
$\mathcal{A}$ be the set of analytic functions $f : \mathcal{U}
\mapsto \mathbb{C}.$  Define the empirical process $G_n : =
\{G_n(f)\}$ indexed by $\mathcal{A}$
\begin{equation}
G_n(f)= p\cdot \int_{-\infty}^{+\infty} f(x)\left[F^{V_n}-
F_{y_{n_1}, y_{n_2}}\right] (dx), \quad f \in \mathcal{A}.
\label{Gdef2}
\end{equation}
Here $F_{y_{n_1}, y_{n_2}}$ is the  distribution in
(\ref{LSDden})  with indexes $y_{n_k}$ (instead of $y_k$), k=1,2.

Recently, \citet{Z08} establishes  a general  CLT for LSS of
large-dimensional F matrix.  The following theorem is a simplified
one quoted from it. {Throughout the paper, $\oint$ denotes a contour
integral along a given contour.}

\begin{theorem}   \label{T2.1}
Let  $ f_1, \cdots ,f_k \in \mathcal{A}$,
and assume:\\
For each p,  $(\xi_{ij_1})$  and $(\eta_{ij_2})$ variables are $ i.i.d. $,
$ 1\leq i\leq p, ~ 1\leq
j_1 \leq n_1, ~  1\leq j_2 \leq n_2. $ $ E\xi_{11}=E\eta_{11}=0, $
$E|\xi_{11}|^4=E|\eta_{11}|^4<\infty,$~ $y_{n_1}=\frac{p}{n_1}
\rightarrow y_1 \in (0, 1),\quad
y_{n_2}=\frac{p}{n_2} \rightarrow y_2 \in (0, 1).$
\begin{itemize}
\item[(i)] Real Case. \quad
  Assume moreover $ (\xi_{ij}) $  and $
  (\eta_{ij}) $ are real, $ E|\xi_{11}|^2=E|\eta_{11}|^2=1 $, then the
  random vector $ \left(G_n(f_1), \cdots,G_n(f_k)\right) $ weakly
  converges to a k-dimensional Gaussian vector  with the mean vector
  \begin{eqnarray}
    m(f_j)&=&
    \lim\limits_{r \rightarrow 1_{+}}
    (\ref{E1})+(\ref{E1betax})+(\ref{E1betay}) \nonumber\\
    &&\quad\frac{1}{4\pi i}\oint_{|\zeta|=1}
    f_j(z(\zeta))\left[\frac{1}{\zeta-{1\over r}}+\frac{1}{\zeta+{1\over
          r}}-\frac{2}{\zeta+{y_2\over
          {h}}}\right]d\zeta \label{E1}\\
    &&\quad + \frac{\beta\cdot y_1(1-y_2)^2}{2\pi i \cdot
      h^2}\oint_{|\zeta|=1}f_j(z(\zeta))\frac{1}{(\zeta+\frac{y_2}{h})^3}d\zeta\label{E1betax}\\
    &&\quad + \frac{\beta\cdot y_2(1-y_2)}{2 \pi i \cdot
      h}\oint_{|\zeta|=1}f_j(z(\zeta))\frac{\zeta +
      \frac{1}{h}}{(\zeta+\frac{y_2}{h})^3}d\zeta ,\label{E1betay}
    \quad\quad j=1, \cdots, k,
  \end{eqnarray}
  where $
  z(\zeta)=(1-y_2)^{-2}\left[1+h^2+2h\mathcal{R}(\zeta)\right], \quad
  h =\sqrt{y_1+y_2-y_1y_2}$,  $ \beta=E|\xi_{11}|^4-3,  $
  and the covariance function
  \begin{eqnarray}
    \lefteqn{\upsilon(f_j, f_\ell)=
      \lim_{r  \rightarrow  1_{+}} (\ref{cov1})
      +(\ref{cov1betax})}  \nonumber\\
    &&  -\displaystyle\frac{1}{2\pi^2}\oint_{|\zeta_2|=1}
    \oint_{|\zeta_1|=1}\frac{f_j(z(\zeta_1))f_\ell(z(\zeta_2))}{(\zeta_1-r\zeta_2)^2}
    d\zeta_1d\zeta_2,\label{cov1}\\
    && -\frac{\beta \cdot (y_1+y_2)(1-y_2)^2}{4\pi^2h^2}
    \oint_{|\zeta_1|=1}\frac{f_j\left( z(\zeta_1)
      \right)}{(\zeta_1+\frac{y_2}{h})^2}d\zeta_1
    \oint_{|\zeta_2|=1}\frac{f_\ell\left( z(\zeta_2)
      \right)}{(\zeta_2+\frac{y_2}{h})^2}d\zeta_2
    \label{cov1betax}\\
    && j, \ell \in \{1,\cdots , k\}.\nonumber
  \end{eqnarray}
\item[(ii)] Complex Case. Assume moreover $ (\xi_{ij}) $  and $
  (\eta_{ij}) $ are complex,  $ E(\xi_{11}^{2})=E(\eta_{11}^{2})=0, $
  then the conclusion of (i)
  also holds, except the means are
  $     (\ref{E1betax})+(\ref{E1betay}) $ and
  the covariance function is
  $ \frac12  \lim\limits_{r \rightarrow 1_{+}}
  (\ref{cov1})  +  (\ref{cov1betax})$ with
  $  \beta=E|\xi_{11}|^4-2 $.
\end{itemize}
\end{theorem}

We should point out that Zheng's CLT for $F$-matrices covers more
general
situations the those cited in Theorem~\ref{T2.1}. In particular, the
fourth moments $E|\xi_{11}|^4 $ and $E|\eta_{11}|^4 $ can be different.

The following lemma will be used in \S\ref{sec:Clinear} for an
application of Theorem \ref{T2.1} (see
 (\ref{testE}) and (\ref{testVar})).
For a proof, see \cite{B2009}.

\begin{lemma}
For the function
 $f(x)=\log(a+bx),\quad x\in \mathbb{R}, \quad a, b>0 $, let
 $(c,d)$ be the unique solution to the equations
 \begin{equation}
 \left\{
 \begin{array}{llll}
c^2+d^2=a+b\frac{(1+h^2)}{(1-y_2)^2},\\
 cd=\frac{bh}{(1-y_2)^2},\\
0 <d< c.\\
\end{array} \right.\label{cd}
\end{equation}
Analogously,  let  $\gamma, \eta$ be the constants similar to $(c,
d)$ but for the function $g(x)=\log(\alpha+\beta x), \quad \alpha
> 0,\quad \beta > 0.$
 Then,  the mean and covariance functions in (\ref{E1}) and
 (\ref{cov1}) equal to
 \begin{eqnarray*}
   m(f)&=&\frac{1}{2}\log\frac{(c^2-d^2)h^2}{(ch-y_2d)^2},\\
   \upsilon(f, g)& =&
   2\log{\frac{c\gamma}{c\gamma-d\eta}}.
 \end{eqnarray*}
 \label{lem1}
\end{lemma}

\section{Testing  a general linear hypothesis in high-dimensional regressions}
\label{sec:Clinear}

\subsection{A corrected LR test}

The construction of a correct scaling for the LRT statistic $\bLa_n$
of the test (\ref{H02B}) will rely on the CLT~\ref{T2.1}. Recall
that
\[   \bLa_n=\left|\textbf{I}+  \frac{q_1}{n-q} \bF \right|^{-1},\quad
\textbf{F}=\frac{n-q}{q_1}(n\widehat{\bSi})^{-1}(\widehat{\textbf{B}}_1-\textbf{B}_1^*)
\textbf{A}_{11:2}(\widehat{\textbf{B}}_1-\textbf{B}_1^*)'.
\]
Under $H_0$, we have
\[
n\widehat{\bSi} \sim W_p(\bSi,n-q),\qquad
(\widehat{\textbf{B}}_1-\textbf{B}_1^*)
\textbf{A}_{11:2}(\widehat{\textbf{B}}_1-\textbf{B}_1^*)' \sim
W_p(\bSi,q_1),
\]
and they are independent.
Consequently,  $\textbf{F}$ is exactly distributed as the $F$-matrix
$V_n$ defined in (\ref{F}), where in addition all the variables are
Gaussian.

Our correction to the LRT statistic $\bLa_n$
is given in the following theorem.

\begin{theorem} \label{T4.1}
  For  the general linear hypothesis (\ref{H02B}) in
  the regression model \eqref{linear model},  let $\bLa_n$ be Wilk's LRT
  statistic  given in    (\ref{lamda*}).
  Define also the function
  \[
  f(x)=\log(1+\frac{y_{n_2}}{y_{n_1}}x)~,
  \]
  and assume that
  \begin{equation}
    \label{eq:pqn}
    p\to\infty,~~
    q_1\to\infty,\quad
    n-q\to\infty,\quad
    y_{n_1}=\frac{p}{q_1}\to y_1 \in (0,1),\quad
    y_{n_2}=\frac{p}{n-q}\to y_2 \in (0,1).
  \end{equation}
  Then, under the null,
  \begin{equation}\label{LST}
    T_n=\upsilon(f)^{-\frac{1}{2}}\left[
      -\displaystyle\log \bLa_n-p \cdot F_{y_{n_1},y_{n_2}}(f)-
      m(f)\right] \Rightarrow \CN \left( 0, 1\right),
  \end{equation}
  where $m(f),\upsilon(f)  $ and $ F_{y_{n_1},y_{n_2}}(f)$ are
  defined in  (\ref{testE})(\ref{testVar})and (\ref{limit}), respectively.
\end{theorem}

Before giving a proof, it is worth mentioning that at a first look,
the asymptotic framework depicted in \eqref{eq:pqn} seems complicated.
Indeed, this is a common set-up in RMT and simply  requires that the degrees
of freedom of the underlying Wishart matrices grow to infinity in a
proportional way with the sample size.

\begin{proof}
  Since   $\textbf{F}$  can be represented by a Gaussian $V_n$, we
  have
  \begin{eqnarray*}
    -\log\bLa_n  &=&\log|I+\frac{q_1}{n-q}V_n|   \\
    &=&\sum\limits_{i=1}^p \log(1+\frac{q_1}{n-q}\lambda^{V_n}_i)\\
    &=& p \cdot \int\log (1+\frac{q_1}{n-q}x)   dF^{V_n}(x).
  \end{eqnarray*}
  Define $f(x)=\log (1+\frac{q_1}{n-q}x)$, by
  $y_{n_1}=p/q_1,y_{n_2}=p/(n-q)$ , also it can be written as
  \begin{equation}
    f(x)=\log(1+\frac{y_{n_2}}{y_{n_1}}x).\label{f(x)}
  \end{equation}
  From
  \begin{eqnarray}
    -\log \bLa_n&=&p \cdot \int f(x)dF^{V_n}(x)\nonumber\\
    &=& p \cdot \int f(x) d\left(F^{V_n}(x)-F_{y_{n_1},
      y_{n_2}}(x)\right) +p \cdot F_{y_{n_1}, y_{n_2}}(f)\nonumber
  \end{eqnarray}
  where $F_{y_{n_1}, y_{n_2}}(f)=\int f(x)dF_{y_{n_1}, y_{n_2}}(x) $
  and $F_{y_{n_1}, y_{n_2}}(x)$ is the limiting distribution which has
  a density in (\ref{LSDden}) but with $y_{n_k}$ instead of $y_k,
  k=1,2.$
  Then we get
  \begin{equation}
    G_n(f)=-\log \bLa_n  -p \cdot F_{y_{n_1}, y_{n_2}}(f).\label{ESD-pLSD}
  \end{equation}

  By Theorem \ref{T2.1}, $ G_n(f)$  weakly converges to a Gaussian
  vector with mean
  \begin{equation}
    m(f)= \frac{1}{2}\log\frac{(c^2-d^2)h^2}{(ch-y_2d)^2}\label{testE}
  \end{equation}
  and variance
  \begin{equation}
    \upsilon(f)=2\log \left(\frac{c^2}{c^2-d^2}\right)\label{testVar}
  \end{equation}
  for the real case, where
  \begin{eqnarray*}
    h&=&\sqrt{y_1+y_2-y_1y_2}\\
    a_0,b_0&=&\frac{(1\mp h)^2}{(1-y_2)^2}\\
    c,d&=&\frac1{2}\left[\sqrt{1+\frac{y_2}{y_1}b_0}\pm
      \sqrt{1+\frac{y_2}{y_1}a_0} \right], c>d.
  \end{eqnarray*}
  This is calculated  in \S\ref{sec:proofs}  using  Lemma~\ref{lem1}.
  For the complex case, the mean $
  m(f) $ is zero and the variance is half of $ \upsilon(f) $. In other
  words,
  \begin{eqnarray}
    -\log\bLa_n-p\cdot F_{y_{n_1}, y_{n_2}}(f)
    &\Rightarrow & N\left(m(f), \upsilon(f)\right).\label{2connec}
  \end{eqnarray}
  Here
  \begin{equation}
    \displaystyle{F_{y_{n_1}, y_{n_2}}(f)} =
    \frac{y_{n_2}-1}{y_{n_2}}\log c_n+
    \frac{y_{n_1}-1}{y_{n_1}}\log(c_n-d_nh_n)+\frac{y_{n_1}+y_{n_2}}{y_{n_1}y_{n_2}}
    \log \left(\frac{c_nh_n-d_ny_{n_2}}{h_n}\right),\label{limit}
  \end{equation}
  where
  \begin{eqnarray*}
    h_n&=&\sqrt{y_{n_1}+y_{n_2}-y_{n_1}y_{n_2}}\\
    a_n,b_n&=&\frac{(1\mp h_n)^2}{(1-y_{n_2})^2}\\
    c_n,d_n&=&\frac1{2}\left[\sqrt{1+\frac{y_{n_2}}{y_{n_1}}b_n}\pm
      \sqrt{1+\frac{y_{n_2}}{y_{n_1}}a_n} \right], c_n>d_n,
  \end{eqnarray*}
  is derived in  \S\ref{sec:proofs} using  the density
  function of $F_{y_{n_1},  y_{n_2}}$.
  Then  we get letting $q_1 \wedge  (n-q_1)\rightarrow\infty$,
  $$
  T_n=\upsilon(f)^{-\frac{1}{2}}\left[ -\displaystyle\log \bLa_n-p
    \cdot  F_{y_{n_1}, y_{n_2}}(f)- m(f)\right] \Rightarrow N \left( 0,
  1\right).
  $$
\end{proof}

We call {\em Corrected likelihood ratio test} (CLRT) for testing
\eqref{H02B}
 the test based on the statistic $T_n$ and its asymptotic
distribution derived in the theorem  above.
Moreover, it is worth noticing that in the above proof, we used the Gaussian
assumption for entry variables to fit $\bF$ to a Gaussian
$F$-matrix. However,
Theorem~\ref{T2.1} does not need this Gaussian assumption. Therefore,
we can expect (or conjecture) that the asymptotic distribution for
$T_n$ in Theorem~\ref{T4.1}, hence the CLRT,  could be valid  more
generally. However, the kurtosis parameter $\beta$ appeared in
Theorem~\ref{T2.1} is no more null and it will appears in the
asymptotic parameters $m(f)$ and $\upsilon(f)$ above.

\subsection{Two least-squares based procedures  for testing \protect\eqref{H02B}}
\label{sec:ST1ST2}

To evaluate the corrected LRT,
we consider two
additional procedures based on least-squares type statistics
as suggested in  \citet{B96}.
We  first need to
find the asymptotic distributions of these statistics.

By (\ref{BMLE}) and the partition of $\textbf{B}$, we obtain
\begin{equation}
  \widehat{\textbf{B}}_1=\sum\limits_{i=1}^n\textbf{x}_i\textbf{z}_{i,1}'\textbf{A}_{11:2}^{-1}
  -\sum\limits_{i=1}^n\textbf{x}_i\textbf{z}_{i,2}'\textbf{A}_{22}^{-1}\textbf{A}_{21}
  \textbf{A}_{11:2}^{-1}.\label{B10MLE}
\end{equation}
Let
\begin{eqnarray}
  M_{n,1}&=&\mbox{tr}
  \left((\widehat{\textbf{B}}_1-\textbf{B}_1^*)(\widehat{\textbf{B}}_1-\textbf{B}_1^*)'\right),\\
  M_{n,2}&=&\mbox{tr}
  \left((\widehat{\textbf{B}}_1-\textbf{B}_1^*)\textbf{A}_{11:2}(\widehat{\textbf{B}}_1-\textbf{B}_1^*)'\right).
\end{eqnarray}
Because $\widehat{\textbf{B}}$ is a unbiased estimator of
$\textbf{B}$, then $E\widehat{\textbf{B}}_1=\textbf{B}_1^*$ under
the null hypothesis. Thus
\begin{eqnarray}
  EM_{n,1}&=& \mbox{tr} (\bSi) \mbox{tr} (\textbf{A}_{11:2}^{-1}),\\
  EM_{n,2}&=& q_1 \mbox{tr} (\bSi),\\
  \sigma^2_{n,1}&=&Var(M_{n,1})=2\mbox{tr}(\bSi^2)\mbox{tr}(\textbf{A}_{11:2}^{-2})+\beta_x\beta_{z1},\\
  \sigma^2_{n,2}&=&Var(M_{n,2})=2q_1\mbox{tr}(\bSi^2)+\beta_x\beta_{z2},
\end{eqnarray}
where
\begin{eqnarray*}
  \beta_x &=& E(\varepsilon_1'\varepsilon_1)^2-(\mbox{tr}(\bSi))^2-2\mbox{tr}(\bSi^2),\\
  \beta_{z1} &=&
  \sum\limits_{i=1}^n\left[(\textbf{z}_{i,1}'-\textbf{z}_{i,2}'\textbf{A}_{22}^{-1}\textbf{A}_{21})
    \textbf{A}_{11:2}^{-2}
    (\textbf{z}_{i,1}-\textbf{A}_{12}\textbf{A}_{22}^{-1}\textbf{z}_{i,2})\right]^2,\\
  \beta_{z2} &=&
  \sum\limits_{i=1}^n\left[(\textbf{z}_{i,1}'-\textbf{z}_{i,2}'\textbf{A}_{22}^{-1}\textbf{A}_{21})
    \textbf{A}_{11:2}^{-1}
    (\textbf{z}_{i,1}-\textbf{A}_{12}\textbf{A}_{22}^{-1}\textbf{z}_{i,2})\right]^2.\\
\end{eqnarray*}
Define
\begin{equation}
  \textbf{Z}_i^{(k)}=\textbf{A}_{11:2}^{-(3-k)/2}
  (\textbf{z}_{i,1}-\textbf{A}_{12}\textbf{A}_{22}^{-1}\textbf{z}_{i,2}),
  \quad k=1,2.
\end{equation}

\begin{theorem}\label{Th.Mnk}
Assuming that
\begin{enumerate}
\item   $\min (q_1,p,n-q) \rightarrow \infty$;
\item As $ p \rightarrow \infty$,
  $\mbox{tr}\bSi^2=\textit{o}((\mbox{tr}\bSi)^2)$;
\item
  $\max\limits_{1\leq i\leq n }
  \textbf{Z}_i^{(k)'}\textbf{Z}_i^{(k)}=o([\mbox{tr}\textbf{A}_{11:2}^{-(2-k)}])$;
\item
  $(\varepsilon_i),~ i=1, \cdots, n$   are   i.i.d.
  zero-mean random vectors such that
  for any   $\eta>0$, there exists a  $K>0$, such that
  \begin{eqnarray*}
    E(\varepsilon_1'\varepsilon_2)^2 &\leq& K(\mbox{tr}\bSi^2),\\
    \max E(\varepsilon_1'\varepsilon_2)^2I\left(|\varepsilon_1'\varepsilon_2|\geq
    \eta\sqrt{\mbox{tr}\textbf{A}_{11:2}^{-2(2-k)}\mbox{tr}\bSi^2}/|\textbf{Z}_i^{(k)'}\textbf{Z}_j^{(k)}|\right)
    &=&\textit{o}(\eta^2(\mbox{tr}\bSi^2)),\\
    E(\varepsilon_1'\varepsilon_1-\mbox{tr}\bSi)^2 &\leq& K(\mbox{tr}\bSi^2),\\
    E(\varepsilon_1'\varepsilon_1-\mbox{tr}\bSi)^2
    I\left(|\varepsilon_1'\varepsilon_1-\mbox{tr}\bSi|\geq
    \eta\sqrt{\beta_{zk}\mbox{tr}\bSi^2}/|\textbf{Z}_i^{(k)'}\textbf{Z}_j^{(k)}|\right)
    &=&\textit{o}(\eta^2(\mbox{tr}\bSi^2)).
  \end{eqnarray*}
\end{enumerate}
Then for  $k=1,2$ and under $H_0$ in \eqref{H02B},
\[
\Gamma_{n,k}:=\frac{M_{n,k}-EM_{n,k}}{\sigma_{n,k}} \Rightarrow \CN(0,1).
\]
\end{theorem}

Consequently, to test \eqref{H02B}, we can use any  of the statistics
$\Gamma_{n,1}$ and
$\Gamma_{n,2}$.
These tests will be referred below  as ST1 and ST2.

\subsection{A simulation study for  comparison of the  tests}
\label{sec:simulinear}

We set up a simulation experiment to compare five procedures
for testing  \eqref{H02B}:
the classical LRT with an asymptotic  $\chi^2$ approximation,
the associated Bartlett-Box correction (BBC) recalled in
\eqref{eq:BBC}, 
our corrected LRT (CLRT) introduced in \S\ref{sec:Clinear}
and the two tests ST1 and ST2 based on least-squares type statistics
of  \S\ref{sec:ST1ST2}.
 Denote the non-center parameter as ~$\psi=c_0^2\psi_0$, where $\psi_0= {\rm tr}
\left((\textbf{B}_1-\textbf{B}_1^*)'
\bSi^{-1}(\textbf{B}_1-\textbf{B}_1^*)\right)$,
and $c_0$ is a varying constant. 
Then we consider the model (\ref{linear model}) as
the form
$\textbf{x}_i=c_0(\textbf{B}_1-\textbf{B}_1^*)\textbf{z}_i+\boldsymbol{\varepsilon}_i,
\quad i=1,\ldots,n.$ Assume that the
elements of ~$(\textbf{B}_1-\textbf{B}_1^*)$ follow the distribution $\CN(1,1)$.
All the
$i.i.d.$ elements of $\textbf{z}_i$ in the model are sampled from
$\CN(1,0.5)$. The errors  $\boldsymbol{\varepsilon}_i$ in (\ref{linear model})
have a multivariate normal distribution $\CN_p(0,C)$  with
\[
C=\left(
\begin{array}{ccccc}
  1 & \rho & \rho^2 & \cdots & \rho^{p-1} \\
  \rho & 1 & \rho & \cdots & \rho^{p-2} \\
  \ldots & \ldots &  &  &  \\
  \rho^{p-1} & \rho^{p-2} & \cdots & \rho & 1 \\
\end{array}
\right).
\]
Therefore, $\rho$ measures the degree of correlations between the
$p$ coordinates of the noise vectors. To understand the effect of
these correlations on the test procedures, we consider two cases:
$\rho=0.9$ and $\rho=0$.

For different values of $(p, n, q, q_1)$, we compute the realized
sizes (Type-I errors) of the five tests   based on 1,000 independent replications.
All the tests are defined with an nominal (and asymptotic)  level
$\alpha= 0.05$. The powers of the tests are evaluated under
alternative hypotheses obtained by varying the parameter $c_0$.

Table~\ref{table00} gives the sizes  (line $c_0=0$, in bold) 
and the powers ($c_0 \neq 0$) for the
case $\rho=0$ and various choices of the dimensions $(p, n, q, q_1)$.
Table~\ref{table09} displays analogous results for the case  $\rho=0.9$
where the coordinates of the noise sequence are highly correlated.
The important conclusions  from these tables are as follows.

\begin{description}
\item[Test size:]
  \begin{itemize}
  \item  The LRT and  BBC correction are highly inconsistent:
    in all considered cases, 
    the LRT and its BBC correction   have a 
    much higher size  than the nominal value  5\%. In
    particular,  
    the LRT  systematically  rejects the null hypothesis,
    even for data dimension as small as $p=10$, while 
    the BBC correction  is just less biased as  expected. 

  \item 
    In the case where the coordinates of the noise are uncorrelated  
    (Table~\ref{table00}), 
    the three tests CLRT, ST1 and ST2 which are  based on the RMT,  
    achieve a correct level   close to 5\%.  
   
    In contrary, when these correlations are high (Table~\ref{table09}),
    as the  least-squares type tests ST1 and   ST2 heavily depend 
    on an assumed non correlation between these  coordinates, these two
    tests become inconsistent. 
  \end{itemize}

\item[The power function:]

  In the case where the coordinates of the noise are uncorrelated  
  (Table~\ref{table00}),  while being all consistent, 
  CLRT and ST2 outperform the test ST1. 

  When  these coordinates are highly  correlated
  (Table~\ref{table09}) and  despite their  inconsistency, 
  the  tests ST1 and ST2 are outperformed by the CLRT. 
  For example, in  the case $\rho=0.9,n=200,p=30$, 
  the highest power of ST1 and  ST2 are only 0.283 and 0.115,
  respectively. 
\end{description}

To summarize, among the five tests considered here, only the CLRT
displays an overall consistency and a generally satisfactory power. In
particular, this test is robust with regard to the correlations
between the coordinates of the noise process. 

Lastly, Figures 1 and 2  give a dynamic view of these
comparisons by varying the non-central parameter $c_0$ for the cases
$\rho=0$ and $\rho=0.9$, respectively. Note that the left-first point
of all lines represent the realized sizes (Type I errors) of the
tests,  and others are the powers. 

\section{A high dimensional multiple sample significance test}
\label{sec:Application}

In this section we consider the following
multiple sample significance test  problem in a MANOVA with
high-dimensional data. For the two sample case,
this problem has been considered by
 \citet{D58} and  \cite{B96}. Here we  treat the general multiple
 sample case.
Consider $q$ Gaussian populations $\CN(\mu^{(i)},\bSi)$ of dimension
$p$, $1\le i\le q $, and for
each population, assume that we have  a sample of size $n_i$:
$\{ \bx^{(i)}_{k},~1\le k\le n_i\}$.
We wish to test the hypothesis
\begin{equation}  
  H_0 : ~{\bmu}^{(1)}  = \cdots =  {\bmu}^{(q)}~. \label{H0q}
\end{equation}
High dimensional here  means that both the number $q$ of the populations
 and the dimension $p$ of the observation vectors are large with
respect to the sample sizes $(n_i)$'s.

Clearly, the observations can be put in the form
\begin{equation}\label{modelQ}
  \bx^{(i)}_k = \bmu^{(i)} + \varepsilon^{(i)}_k ,
  \quad 1\le i\leq q,~1\le  k\le n_i,
\end{equation}
where $\{\varepsilon^{(i)}_k\}$ is an array of i.i.d. random vectors
distributed as $\CN_p(0,\bSi)$.
We are going to embed the test \eqref{H0q} into  a special instance of
the regression test \eqref{H02B}.
To this end, let $\{e_i\}$ be the canonical base of $\mathbb{R}^p$ and
we  define the following regression vectors
\[  \bz^{(i)}_k = [e_i + e_q] \mathbf{1}_{\{i<q\}} + e_q
\mathbf{1}_{\{i=q\}}, \quad   1\le i\leq q,~1\le  k\le n_i.
\]
Define moreover the $p\times q$ matrix $\bB=(\bB_1,\bB_2)$ with
\begin{eqnarray}
  \bB_1  &=&(\bmu^{(1)}- \bmu^{(q)}, \ldots, \bmu^{(q-1)}- \bmu^{(q)}),\\
  \bB_2  &=&  {\bmu}^{(q)} .
\end{eqnarray}
Note that the dimension $q$ is  split to $(q_1,q_2)=(q-1,1)$
in the above decomposition.

Therefore, the observations follow a linear model
\begin{equation}\label{modelQ1}
  \bx^{(i)}_k =  \bB \bz^{(i)}_k   + \varepsilon^{(i)}_k ,
  \quad 1\le i\leq q,~1\le  k\le n_i.
\end{equation}
The multiple sample test \eqref{H0q} is equivalent to the following
regression test
\begin{equation}
  H_0 : ~ \bB_1 = 0 ~. \label{H0qR}
\end{equation}

In order to apply Theorem~\ref{T4.1}, we now identify the likelihood
ratio  statistic
$\bLa_n$ defined in \eqref{eq:bLa}. Here denote $n=\sum_{i=1}^{q} n_i$.
Under the null hypothesis, the likelihood estimates of $(\bB_2,\bSi)$
are {(see \cite{A03} for details of computation)}
\begin{eqnarray}
  \widehat{\textbf{B}}_{20} & = & \overline{\textbf{x}} =
  \frac{1}{n}\sum\limits_{i,k}\textbf{x}_k^{(i)}, \\
  \widehat{\bSi}_0 &=&
  \frac{1}{n}\sum\limits_{i,k}(\textbf{x}_k^{(i)}-\overline{\textbf{x}})
  (\textbf{x}_k^{(i)}-\overline{\textbf{x}})'.  \label{sigma0}
\end{eqnarray}
On the other hand, under the alternative hypothesis,
the likelihood estimates of $(\bmu^{(i)},\bSi)$ are
\begin{eqnarray}
  \widehat\bmu^{(i)} & = & \overline{\textbf{x}}^{(i)} :=
  \frac{1}{n_i}\sum\limits_{k=1}^{n_i} \bx_k^{(i)}, \quad 1\le i\le q,\\
  \widehat{\bSi} &=&
  \frac{1}{n}\sum\limits_{i,k}(\textbf{x}_k^{(i)}-\overline{\bx}_k^{(i)})
  (\textbf{x}_k^{(i)}-\overline{\textbf{x}}_k^{(i)})'.  \label{sigma1}
\end{eqnarray}
The likelihood ratio statistic $\bLa_n=|\widehat\bSi|/|\widehat\bSi_0|$ readily
follows.

By application of Theorem~\ref{T4.1}, we have the following
\begin{proposition}\label{prop:M}
  For the multiple sample significance test \eqref{H0q}, assume that
  $q\to\infty$, $n_i\to\infty$, $1\le i\le q$, $p\to\infty$ in such a manner that
  \begin{equation}
    y_{n_1}:=\frac{p}{q-1}  \rightarrow y_1\in(0,1), \qquad
    y_{n_2}=\frac{p}{n-q} \rightarrow y_2\in(0,1).
    \label{y1y2}
  \end{equation}
  Then, for  the same function $f$ defined in (\ref{f(x)}), we have
  \[
  T_n^*=\upsilon(f)^{-\frac{1}{2}}\left[
    -\displaystyle\log \bLa_n-p \cdot F_{y_{n_1},y_{n_2}}(f)-
    m(f)\right] \Rightarrow \CN \left( 0, 1\right).
  \]
  where $\upsilon(f), m(f) $ and $ F_{y_{n_1},y_{n_2}}(f)$ are
  defined in  (\ref{testE}), (\ref{testVar}) and (\ref{limit})
  respectively,
  with the values of  $y_{n_1},y_{n_2}, y_1, y_2$
  defined in  (\ref{y1y2}).
\end{proposition}

It is worth noticing here that the classical likelihood ratio test
(LRT) for testing \eqref{H0q} will rely on the
following weak convergence theorem:
under  $H_0$ and assuming fixed  $p$ and $q$  while letting
$n_i\to\infty$,
\begin{equation}
  -n  \log \bLa_n \Rightarrow \chi^2_{p(q-1)}~.
\end{equation}
Inevitably, in high dimensional case,  $U_n$ will drifts to infinity
by Proposition~\ref{prop:M}. Consequently,
this  classical $\chi^2$-approximation
will leads to a test size much higher than a given
nominal test level, exactly as for  the general linear hypothesis
considered in \S\ref{sec:Clinear}.

\section{Proofs}
\label{sec:proofs}

\subsection*{Proof of (\ref{testE}) and (\ref{testVar}):}
Because $\textbf{x}_i$   are Gaussian variables, for real case,
$\beta=E|$\textbf{$\xi$}$|^4-3=0,$ then (\ref{E1betax}),
(\ref{E1betay}) and (\ref{cov1betax})
 are all 0. Consider (\ref{E1}) and (\ref{cov1}),
as $y_{n_k} \rightarrow y_k, ~k=1,2,$   and during  the process of
Lemma \ref{lem1} calculation, we will see that the constant and
items approaching to zero do not effect on the the circle
integration results, and in practice $ y_{n_k}=y_k, k=1,2. $  So we
use
$$
f(x)=\log(1+\frac{y_2}{y_1}x)
$$
instead of $f(x)=\log(1+\frac{y_{n_2}}{y_{n_1}}x)$.  Make substitute
$x=(1-y_2)^{-2}(1+h^2-2h\cos \theta)$, where
$z(\xi)=(1-y_2)^{-2}\left[1+h^2+2h\mathcal{R}(\xi)\right], \quad h
=\sqrt{y_1+y_2-y_1y_2}$. Because
$$
\log(1+\frac{y_2}{y_1}z(\xi)) =
\log\left(\left|c+d\xi\right|^2\right)
$$
where
\begin{eqnarray*}
c,d&=&\frac1{2}\left[\sqrt{1+\frac{y_2}{y_1}b_0}\pm
\sqrt{1+\frac{y_2}{y_1}a_0} \right], c>d,\\
a_0,b_0&=&\frac{(1+h)^2}{(1-y_2)^2}.
 \end{eqnarray*} is the solution
of the equation (\ref{cd}) with $a,\alpha=1,
b,\beta=\frac{y_2}{y_1}.$
 Then use Lemma \ref{lem1}, we have
\begin{eqnarray*}
m(f)&=& \frac{1}{2}\log\frac{(c^2-d^2)h^2}{(ch-y_2d)^2},\\
\upsilon(f)&=&2\log \left(\frac{c^2}{c^2-d^2}\right)
\end{eqnarray*}
for the real case.

\subsection*{Proof of $F_{y_{n_1}, y_{n_2}}(f)$, Eq.~(\protect\ref{limit}):}

For this computation we drop the indexes  $n_1$ and $n_2$ in the
parameters $y_{n_j}$  and compute the integral
$F_{y_{1}, y_{2}}(f)$.
Following a device designed in \cite{Z08} (Lemma A.2),
let $\underline m(z)$ be the Stieltjes transform of the distribution
function
$ \underline F := (1-y_1)I_{(0,\infty )} + y_1 F_{y_{1}, y_{2}}.
$
For $r>1$ but very close to $1$ and  $|\xi|=1$,  we use a change of
variable  $z=\phi(\xi)$ which is implicitly defined by the formula
$m_0(z)=-(1+hr\xi)/  (1-y_2)$ and we have the following relations
$$z=-\frac{m_0(z)(m_0(z)+1-y_1)}{\left(m_0(z)+\frac{1}{1-y_2}\right)(1-y_2)}\quad\mbox{and}\quad
\underline{m}(z)=\frac{(1-y_2)\left(m_0(z)+\frac{1}{1-y_2}\right)}{m_0(z)(m_0(z)+1)}.$$
Or equivalently,
$$z=\frac{1+h^2+hr^{-1}\bar\xi+hr\xi}{(1-y_2)^2}
\quad\mbox{and}\quad
\underline{m}(z)=\frac{-(1-y_2)^2\xi}{hr(\xi+\frac{1}{hr})(\xi+\frac{y_2}{hr})}.$$
This shows that when $\xi$ anticlockwise runs along the unit circle,
$z$ anticlockwise runs a contour which closely encloses the interval
$\left[a, b\right]$ when $r$ is close to 1 where
$a=\frac{(1-h)^2}{(1-y_2)^2}$ and $b=\frac{(1+h)^2}{(1-y_2)^2}$.  So
we obtain
\begin{eqnarray*}
 F^{y_{1},y_{2}}(f)&=&\int\limits_{a}^{b}f(x)\frac{(1-y_{2})\sqrt{(b-x)(x-a)}}{2\pi
x(y_{1}+y_{2}x)}dx=y_{1}^{-1}\int\limits_{a}^{b}f(x)d\underline F(x)\\
&=&-\frac{1}{2\pi iy_{1}}\oint_{\cal C}f(z)\underline
m(z)dz~~(\mbox{Any contour}~{\cal C}~~\mbox{enclosing the interval}~\left[a, b\right])\\
&=&\frac{1}{2\pi iy_{1}}\oint_{|\xi|=1}\log|c+d\xi|^2\frac{\xi^2-1}{\xi(\xi+\frac1{h})(\xi+\frac{y_{2}}{h})}d\xi\\
&=&\frac{1}{2\pi
iy_{1}}\oint_{|\xi|=1}\left(\log(c+d\xi)+\log(c+d\xi^{-1})\right)\frac{\xi^2-1}{\xi(\xi+\frac1{h})
(\xi+\frac{y_{2}}{h})}d\xi\\
&&~(\mbox{making } \xi^{-1}\to\xi \mbox{ in the second integral })\\
&=&\frac{1}{2\pi iy_{1}}\oint_{|\xi|=1}\log(c+d\xi)
\left(\frac{\xi^2-1}{\xi(\xi+\frac1{h})(\xi+\frac{y_{2}}{h})}
-\frac{h^2}{y_{2}}\frac{\xi^2-1}{\xi(\xi+h)(\xi+\frac{h}{y_{2}})}\right)d\xi\\
&=&\frac{y_{2}-1}{y_{2}}\log(c)+\frac{y_{1}-1}{y_{1}}\log(c-dh)+\frac{y_{1}+y_{2}}{y_{1}y_{2}}\log\left(\frac{ch-dy_2}{h}\right)
\end{eqnarray*}
 where
 \[
 c,~d=\frac{1}{2}\left(
 \sqrt{1+\frac{y_2}{y_1}b} \pm
 \sqrt{1+\frac{y_2}{y_1}a}
 \right),  ~~ c>d.
\]

\pagebreak

\begin{table}[hp]
  \begin{center}

\begin{tabular}{l|@{\quad}rrrrr|@{\quad}rrrrr}
  $\rho=0$  & \multicolumn{5}{c}{\mbox{\hskip-5mm}$(p,n,q,q_1)=(10,100,50,30)$}&
  \multicolumn{5}{c}{$(p,n,q,q_1)=(20,100,60,50)$}
  \\  \hline\hline\\
  Parameter $c_0$& {LRT}&{CLRT}&{BBC}&{ST1}&{ST2}& {LRT}&{CLRT}&{BBC} &{ST1}&{ST2} \\
  \hline\\
  0     &\bf 1 &\bf 0.056 &\bf 0.101 &\bf 0.070 &\bf 0.086  &\bf 1 &\bf 0.047 &\bf 0.672&\bf 0.042 &\bf 0.072\\[1mm]
 $0.01$ &1 &0.064 &0.113 &0.071 &0.096  &1 &0.084 &0.741&0.044 &0.129\\[1mm]
 $0.02$ &1 &0.083 &0.150 &0.080 &0.136  &1 &0.203 &0.879&0.050 &0.395\\[1mm]
 $0.03$ &1 &0.150 &0.224 &0.098 &0.222  &1 &0.381 &0.963&0.063 &0.851\\[1mm]
 $0.04$ &1 &0.247 &0.342 &0.125 &0.387  &1 &0.583 &0.992&0.091 &0.998\\[1mm]
 $0.05$ &1 &0.382 &0.500 &0.156 &0.588  &1 &0.784 &0.999&0.127 &1\\[1mm]
 $0.06$ &1 &0.574 &0.676 &0.200 &0.792  &1 &0.914 &1    &0.173 &1\\[1mm]
 $0.07$ &1 &0.747 &0.829 &0.279 &0.932  &1 &0.979 &1    &0.257 &1\\[1mm]
 $0.08$ &1 &0.885 &0.925 &0.375 &0.988  &1 &0.996 &1    &0.374 &1\\[1mm]
 $0.09$ &1 &0.953 &0.980 &0.496 &0.997  &1 &0.999 &1    &0.526 &1\\[1mm]
 $0.10$ &1 &0.986 &0.990 &0.624 &1      &1 &1     &1    &0.681 &1\\[1mm]
  \hline
\end{tabular}

\vskip 4mm

\begin{tabular}{l|@{\quad}rrrrr|@{\quad}rrrrr}
  $\rho=0$  & \multicolumn{5}{c}{\mbox{\hskip-5mm}$(p,n,q,q_1)=(30,200,80,60)$}&
  \multicolumn{5}{c}{$(p,n,q,q_1)=(50,200,80,70)$} \\
  \hline\hline\\
   Parameter $c_0$& {LRT}&{CLRT}&{BBC}&{ST1}&{ST2}& {LRT}&{CLRT}&{BBC} &{ST1}&{ST2} \\
  \hline\\
  0       &\bf  1     &\bf 0.060 &\bf 0.178  &\bf 0.054 &\bf 0.062 &\bf  1     &\bf 0.056 &\bf 0.495&\bf 0.036 &\bf 0.048 \\[1mm]
  $0.003$ & 1     &0.062 &0.190  &0.055 &0.065 & 1     &0.063 &0.551&0.040 &0.065\\[1mm]
  $0.006$ & 1     &0.078 &0.221  &0.060 &0.083 & 1     &0.099 &0.668&0.042 &0.135\\[1mm]
  $0.009$ & 1     &0.106 &0.276  &0.068 &0.123 & 1     &0.210 &0.797&0.048 &0.372\\[1mm]
  $0.012$ & 1     &0.164 &0.357  &0.071 &0.229 & 1     &0.363 &0.908&0.060 &0.734\\[1mm]
  $0.015$ & 1     &0.232 &0.462  &0.082 &0.352 & 1     &0.560 &0.972&0.073 &0.974\\[1mm]
  $0.018$ & 1     &0.348 &0.584  &0.097 &0.501 & 1     &0.742 &0.991&0.103 &0.999\\[1mm]
  $0.021$ & 1     &0.483 &0.725  &0.131 &0.715 & 1     &0.871 &0.998&0.152 &1\\[1mm]
  $0.024$ & 1     &0.616 &0.831  &0.182 &0.874 & 1     &0.939 &0.999&0.207 &1\\[1mm]
  $0.027$ & 1     &0.771 &0.911  &0.241 &0.970 & 1     &0.984 &1    &0.304 &1\\[1mm]
  $0.03$  & 1     &0.872 &0.954  &0.325 &0.993 & 1     &0.995 &1    &0.414 &1\\[1mm]
  \hline
\end{tabular}
  \end{center}
  \caption{Sizes ($c_0=0$) and powers ($c_0\ne 0$) of  the four methods, based on 1,000 independent applications with real Gaussian
    variables. The parameter $\rho$ in the covariance matrix of errors equals to 0.\label{table00}}
\end{table}

\begin{table}[hp]
  \begin{center}
    \begin{tabular}{l|@{\quad}rrrrr|@{\quad}rrrrr}
      $\rho=0.9$  & \multicolumn{5}{c}{\mbox{\hskip-5mm}$(p,n,q,q_1)=(10,100,50,30)$}&
      \multicolumn{5}{c}{$(p,n,q,q_1)=(20,100,60,50)$} \\
      \hline\hline\\
      Parameter $c_0$& {LRT}&{CLRT}&{BBC}&{ST1}&{ST2}& {LRT}&{CLRT}&{BBC} &{ST1}&{ST2} \\
  \hline\\
        0      &\bf 1 &\bf 0.056 &\bf 0.089 &\bf 0.105 &\bf 0.119  &\bf  1     &\bf 0.055 &\bf 0.681 &\bf 0.087 &\bf 0.155\\[1mm]
       $0.005$ &1 &0.063 &0.099 &0.106 &0.121  & 1     &0.063 &0.696 &0.088 &0.164\\[1mm]
       $0.010$ &1 &0.078 &0.123 &0.107 &0.124  & 1     &0.089 &0.762 &0.089 &0.187\\[1mm]
       $0.015$ &1 &0.110 &0.162 &0.109 &0.134  & 1     &0.165 &0.849 &0.091 &0.220\\[1mm]
       $0.020$ &1 &0.164 &0.234 &0.111 &0.143  & 1     &0.261 &0.923 &0.093 &0.261\\[1mm]
       $0.025$ &1 &0.253 &0.355 &0.116 &0.161  & 1     &0.458 &0.974 &0.095 &0.323\\[1mm]
       $0.030$ &1 &0.388 &0.491 &0.118 &0.182  & 1     &0.690 &0.999 &0.099 &0.408\\[1mm]
       $0.035$ &1 &0.562 &0.652 &0.123 &0.215  & 1     &0.878 &1     &0.101 &0.503\\[1mm]
       $0.040$ &1 &0.724 &0.811 &0.130 &0.250  & 1     &0.963 &1     &0.105 &0.610\\[1mm]
       $0.045$ &1 &0.873 &0.926 &0.136 &0.284  & 1     &0.998 &1     &0.110 &0.704\\[1mm]
       $0.050$ &1 &0.951 &0.979 &0.144 &0.343  & 1     &1     &1     &0.115 &0.801\\[1mm]
      \hline
    \end{tabular}
    \\[5mm]
    \begin{tabular}{l|@{\quad}rrrrr|@{\quad}rrrrr}
      $\rho=0.9$  & \multicolumn{5}{c}{\mbox{\hskip-5mm}$(p,n,q,q_1)=(30,200,80,60)$}&
      \multicolumn{5}{c}{$(p,n,q,q_1)=(50,200,80,70)$} \\
      \hline\hline\\
      Parameter $c_0$& {LRT}&{CLRT}&{BBC}&{ST1}&{ST2}& {LRT}&{CLRT}&{BBC} &{ST1}&{ST2} \\
  \hline\\  0  &\bf  1     &\bf 0.054 &\bf 0.181  &\bf 0.089 &\bf 0.105&\bf  1     &\bf 0.059 &\bf 0.520&\bf 0.098 &\bf 0.100\\[1mm]
       $0.002$ & 1     &0.059 &0.197  &0.090 &0.106& 1     &0.060 &0.536&0.099 &0.107\\[1mm]
       $0.004$ & 1     &0.074 &0.223  &0.090 &0.109& 1     &0.079 &0.604&0.100 &0.116\\[1mm]
       $0.006$ & 1     &0.113 &0.288  &0.091 &0.115& 1     &0.140 &0.697&0.101 &0.136\\[1mm]
       $0.008$ & 1     &0.178 &0.400  &0.091 &0.126& 1     &0.233 &0.811&0.102 &0.175\\[1mm]
       $0.010$ & 1     &0.287 &0.530  &0.092 &0.140& 1     &0.409 &0.913&0.104 &0.230\\[1mm]
       $0.012$ & 1     &0.445 &0.691  &0.093 &0.161& 1     &0.633 &0.979&0.107 &0.300\\[1mm]
       $0.014$ & 1     &0.643 &0.840  &0.097 &0.180& 1     &0.826 &0.993&0.114 &0.379\\[1mm]
       $0.016$ & 1     &0.821 &0.939  &0.101 &0.202& 1     &0.953 &1    &0.118 &0.481\\[1mm]
       $0.018$ & 1     &0.937 &0.986  &0.107 &0.238& 1     &0.992 &1    &0.125 &0.597\\[1mm]
       $0.020$ & 1     &0.987 &0.996  &0.115 &0.283& 1     &1     &1    &0.131 &0.694\\[1mm]
      \hline
    \end{tabular}
  \end{center}
  \caption{Sizes ($c_0=0$) and powers ($c_0\ne 0$) of  the four methods, based on 1,000 independent applications with real Gaussian
    variables. The parameter $\rho$ in the covariance matrix of errors equals to 0.9. \label{table09}}
\end{table}
\begin{figure}[hb]
  \begin{center}
    \includegraphics[width=7cm]{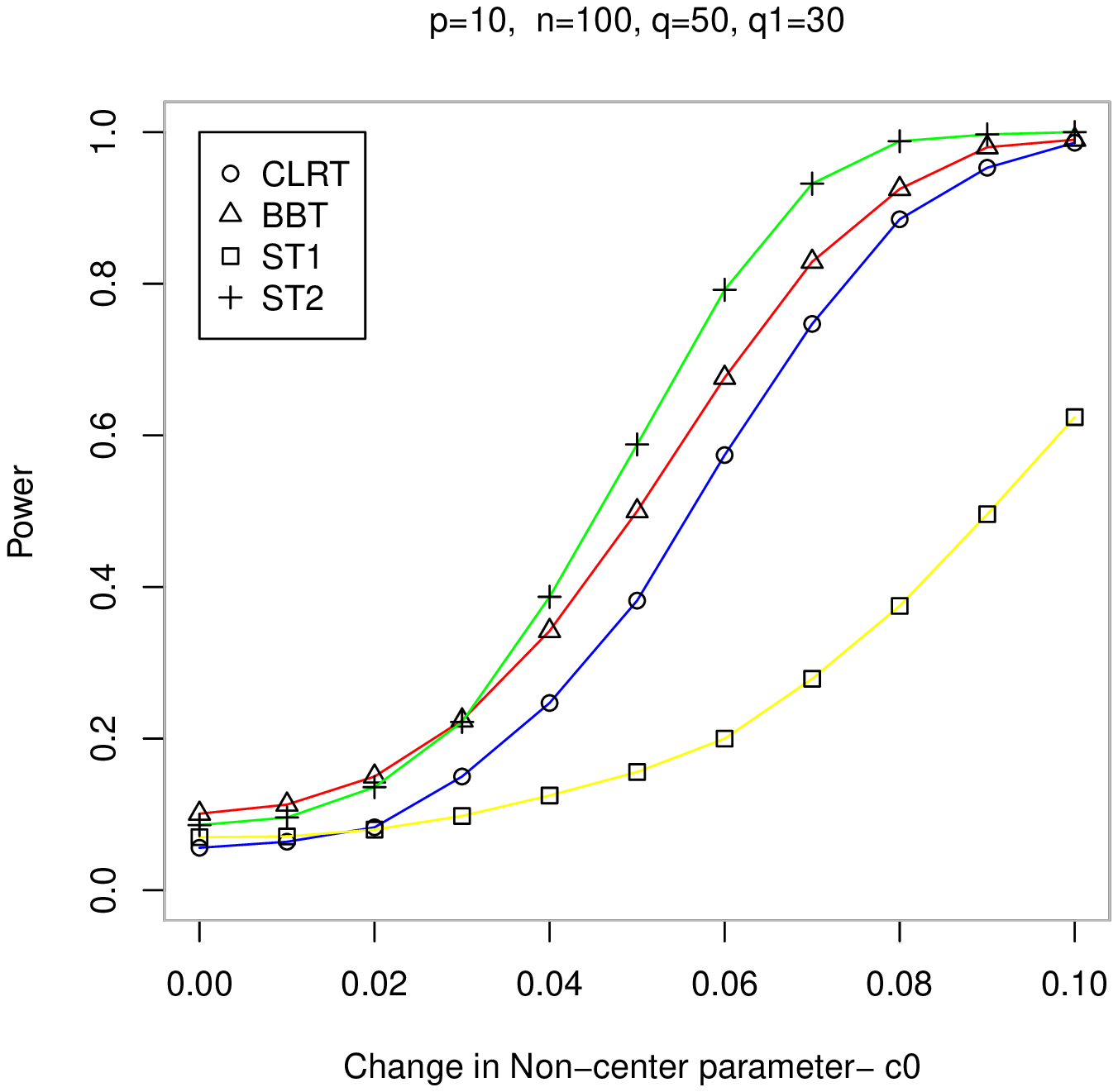} \quad    \includegraphics[width=7cm]{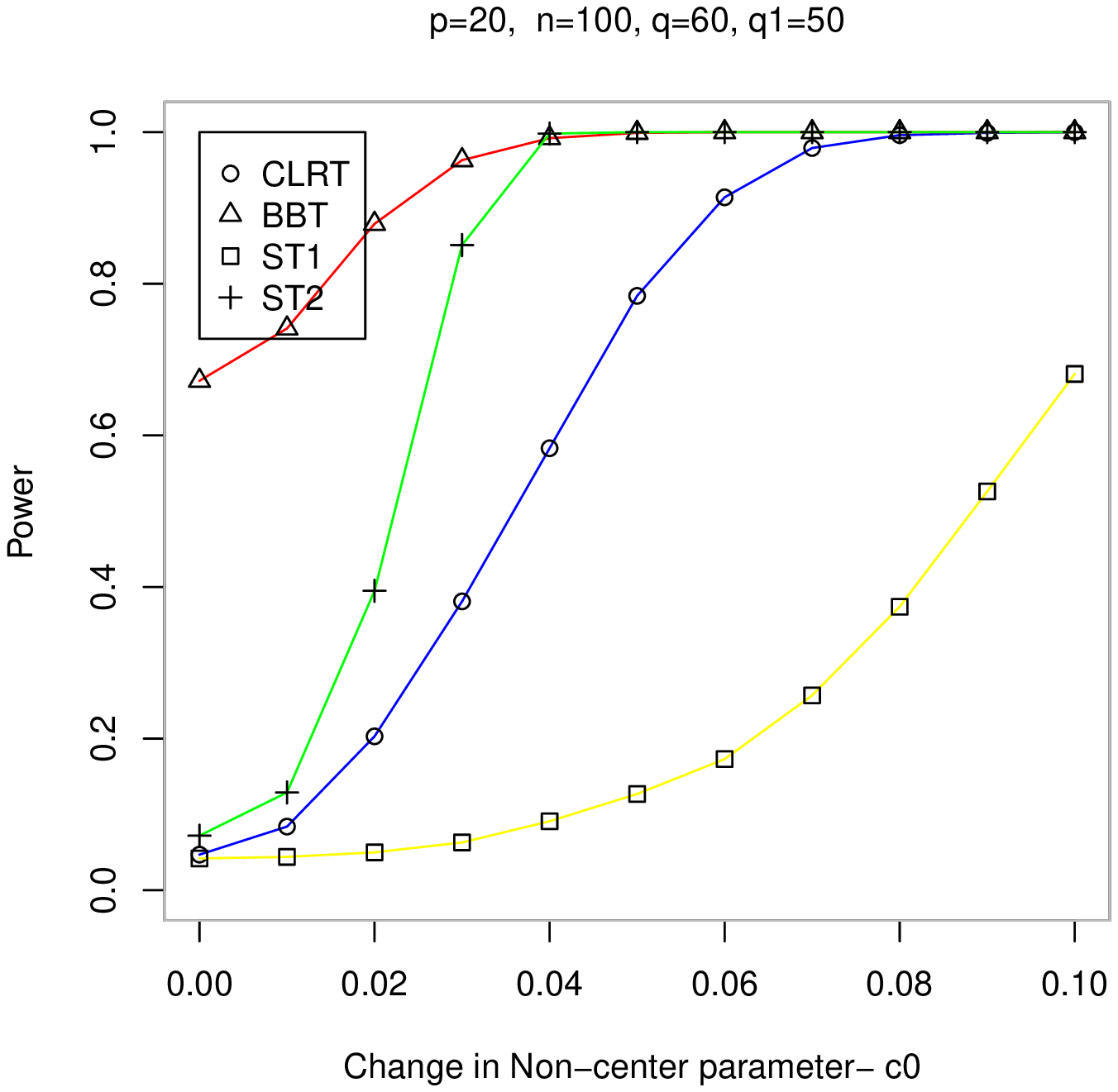}\\[2mm]
    \includegraphics[width=7cm]{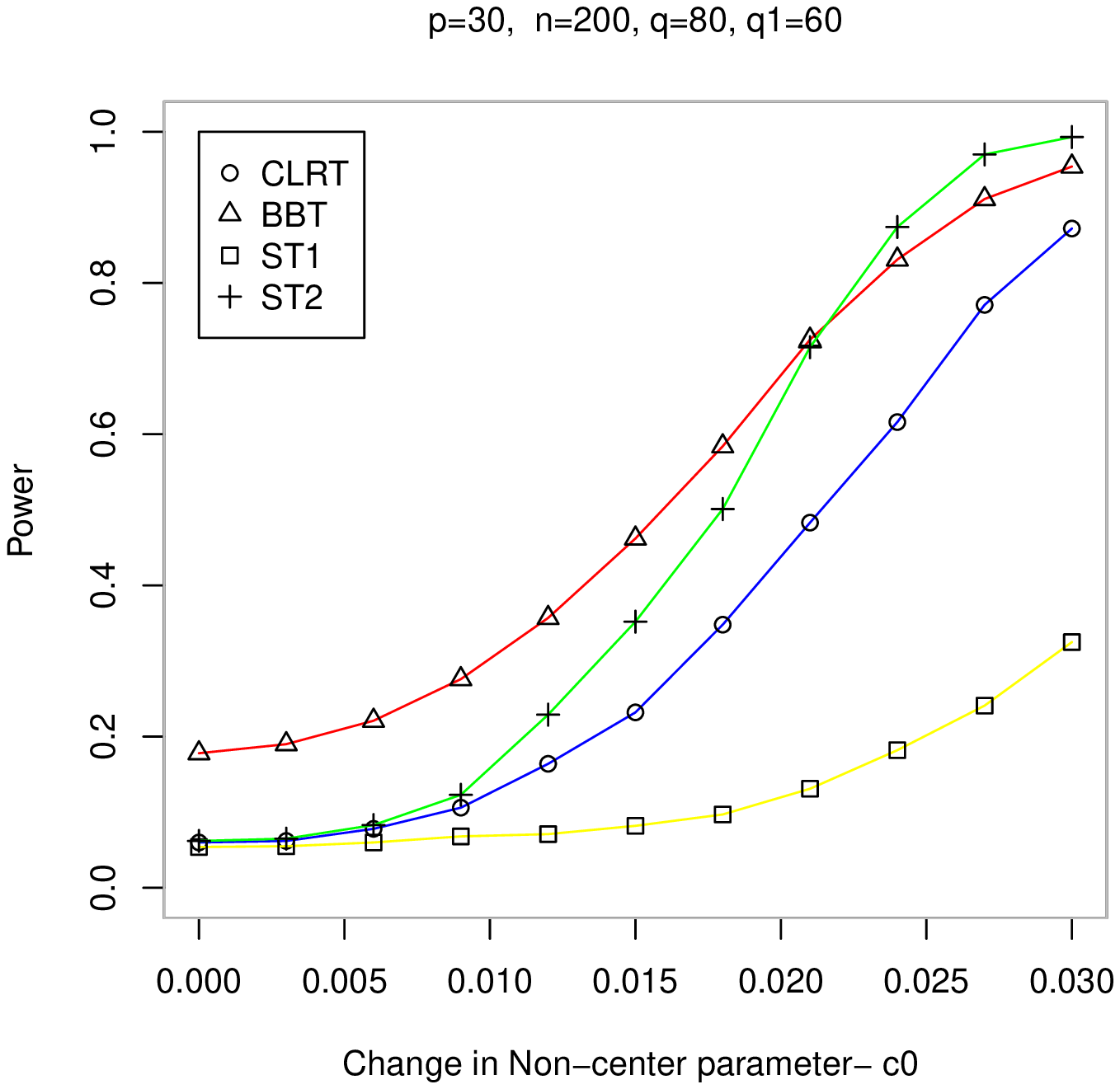} \quad    \includegraphics[width=7cm]{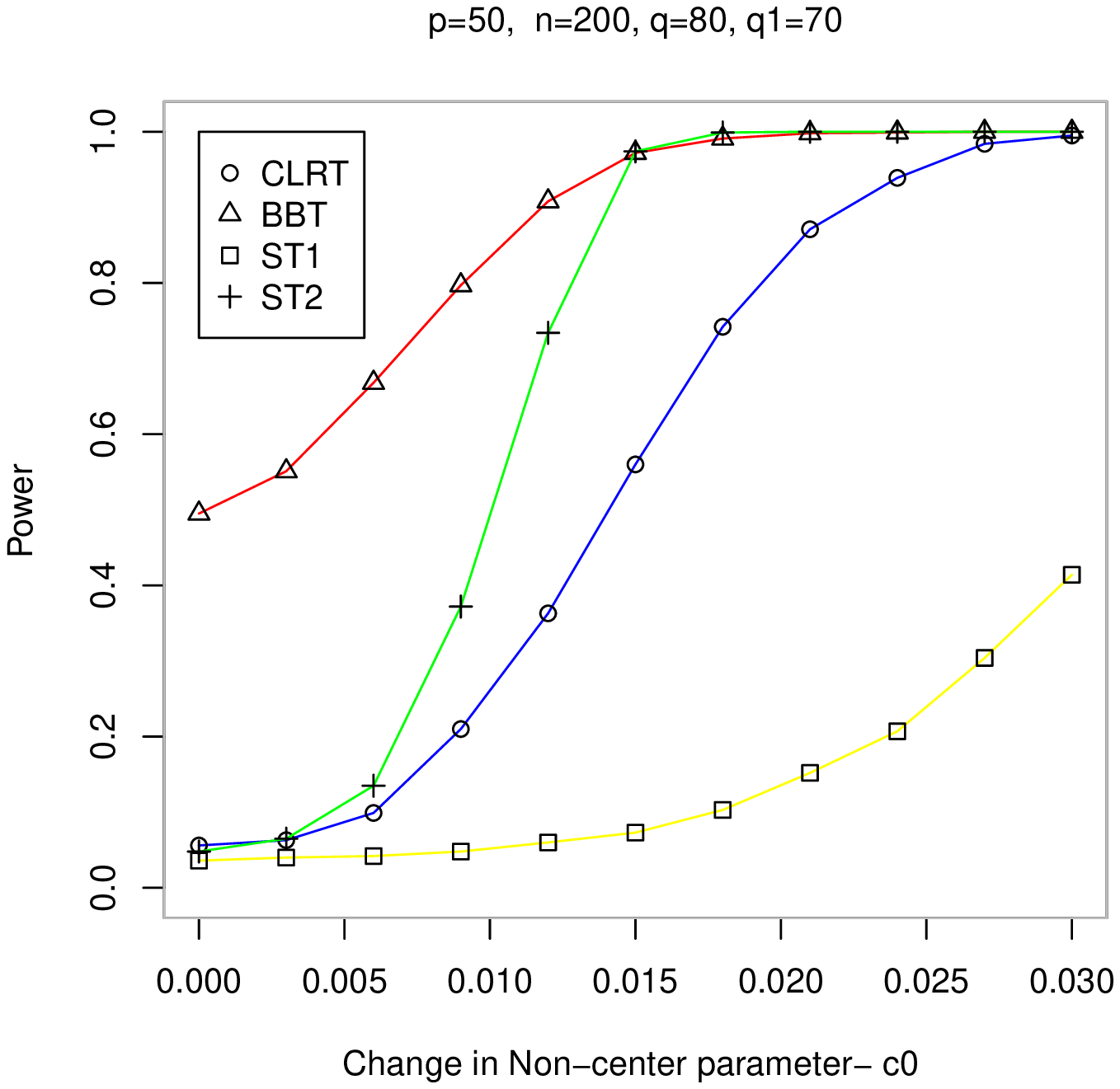}
    \caption{Sizes ($c_0=0$) and Powers ($c_0\ne 0$) of  the four methods, which are the corrected
      LRT (CLRT),  Bartlett-Box correction (BBC) and two least-squares type  tests ( ST1 and ST2 ),  based on 1,000 independent replications using
      Gaussian error variables from $\CN(0,I)$.\quad
      Top row:  ~~   $(p,n,q,q_1)=(10,100,50,30)$ and $(20,100,60,50)$.\quad
      Bottom row:~~$(p,n,q,q_1)=(30,200,80,60)$ and $(50,200,80,70)$.
      \label{fig:12}
    }
  \end{center}
\end{figure}

\begin{figure}[hb]
  \begin{center}
    \includegraphics[width=7cm]{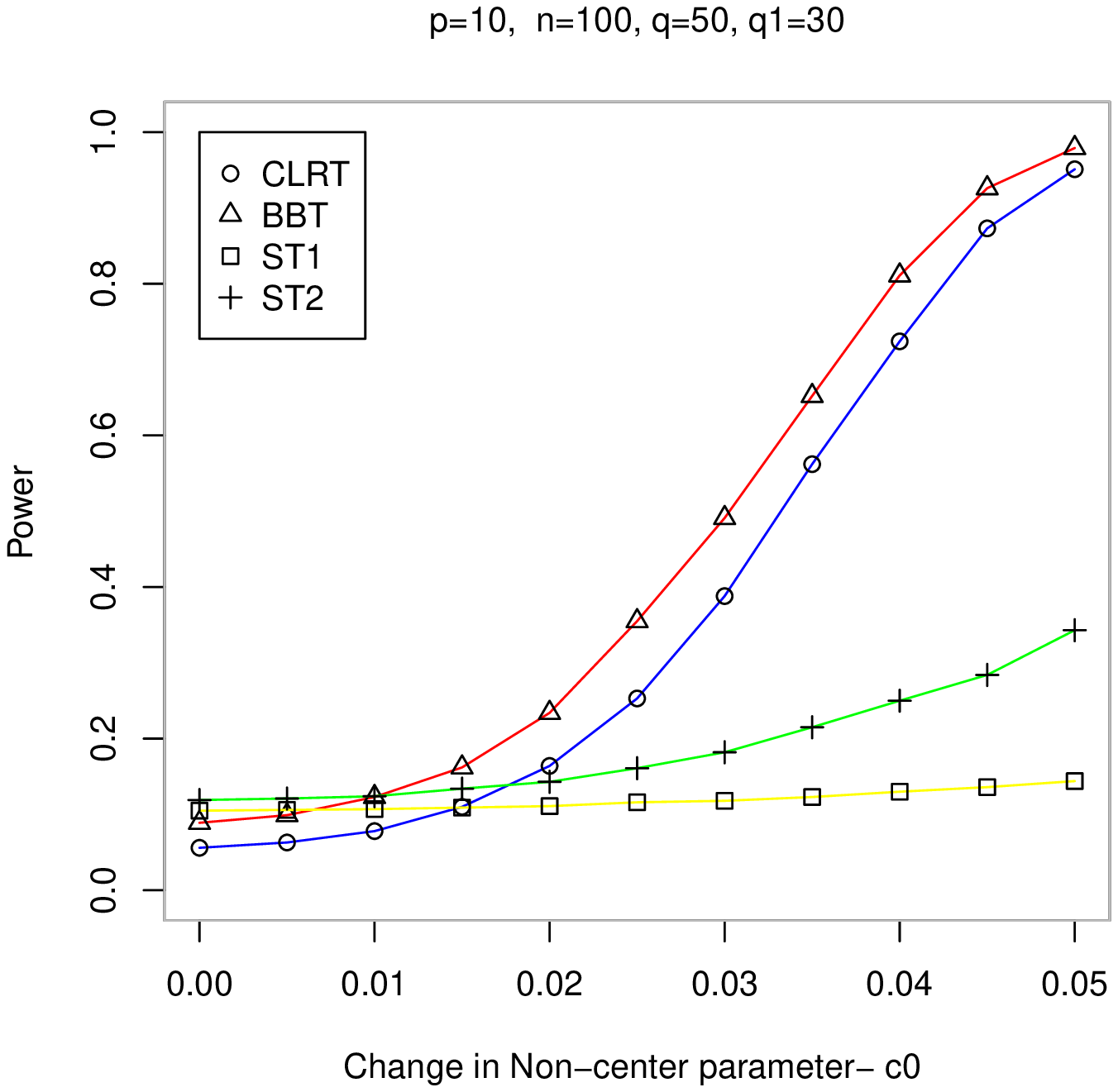}\quad    \includegraphics[width=7cm]{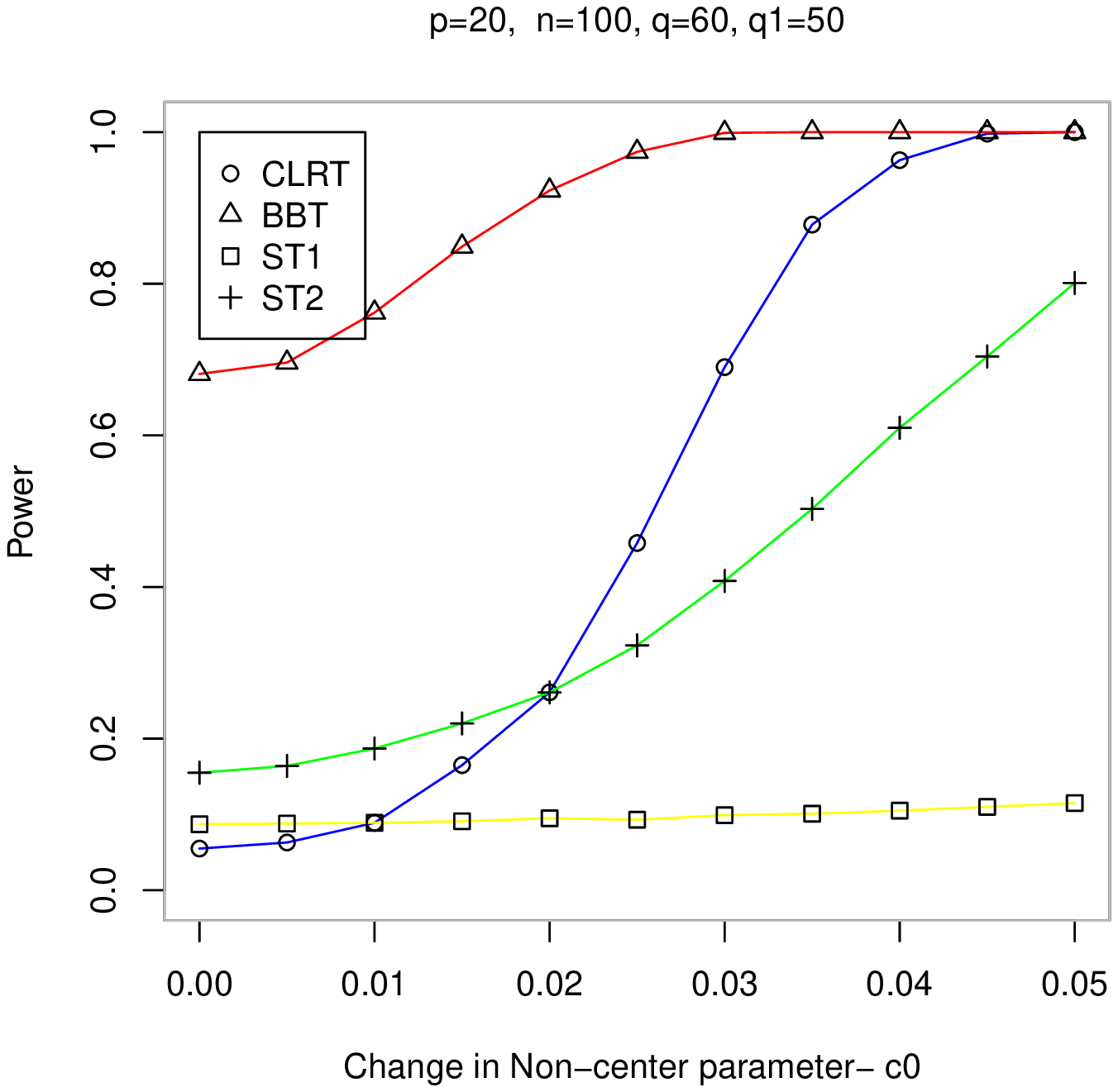}\\[2mm]
    \includegraphics[width=7cm]{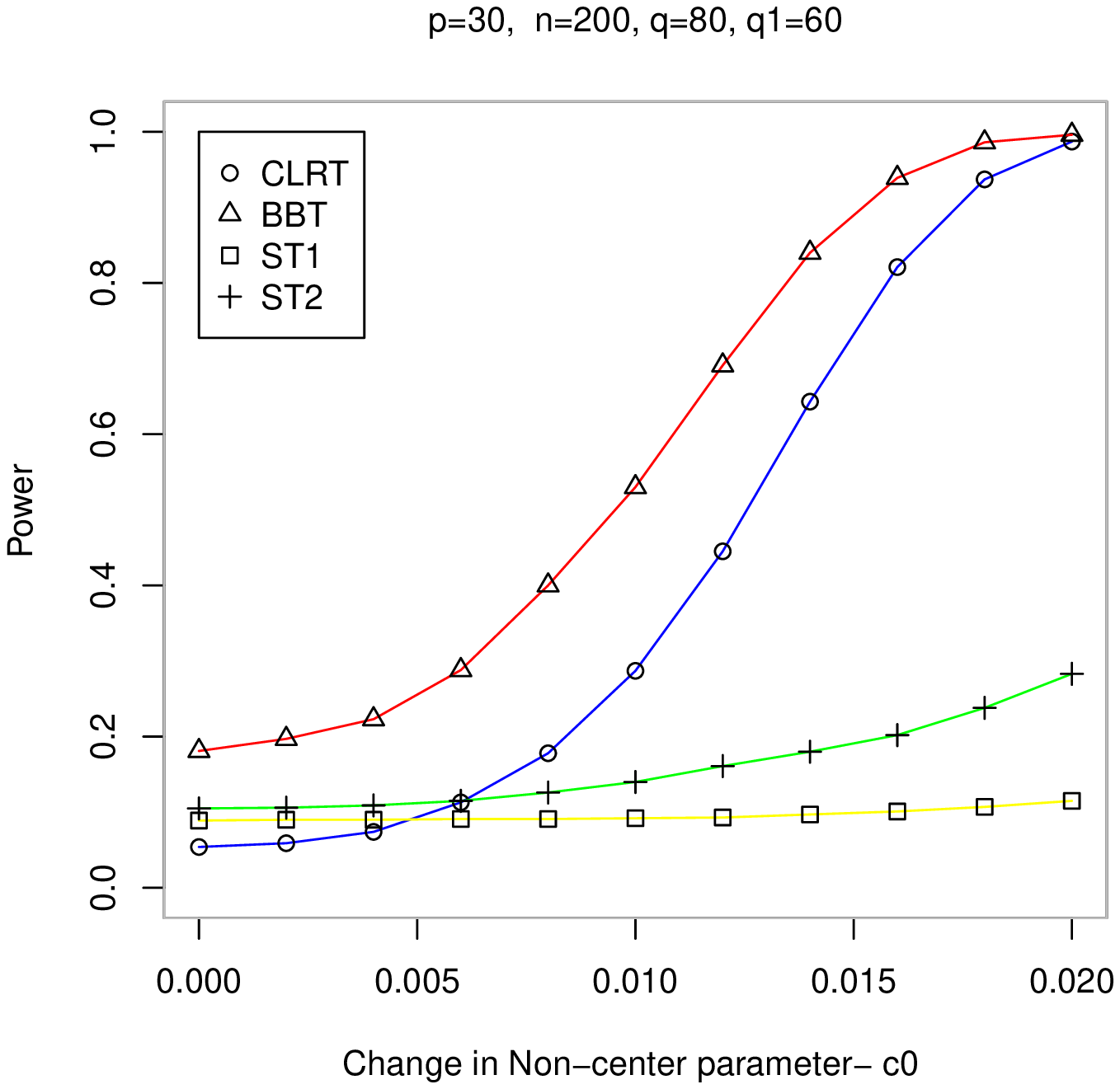}\quad    \includegraphics[width=7cm]{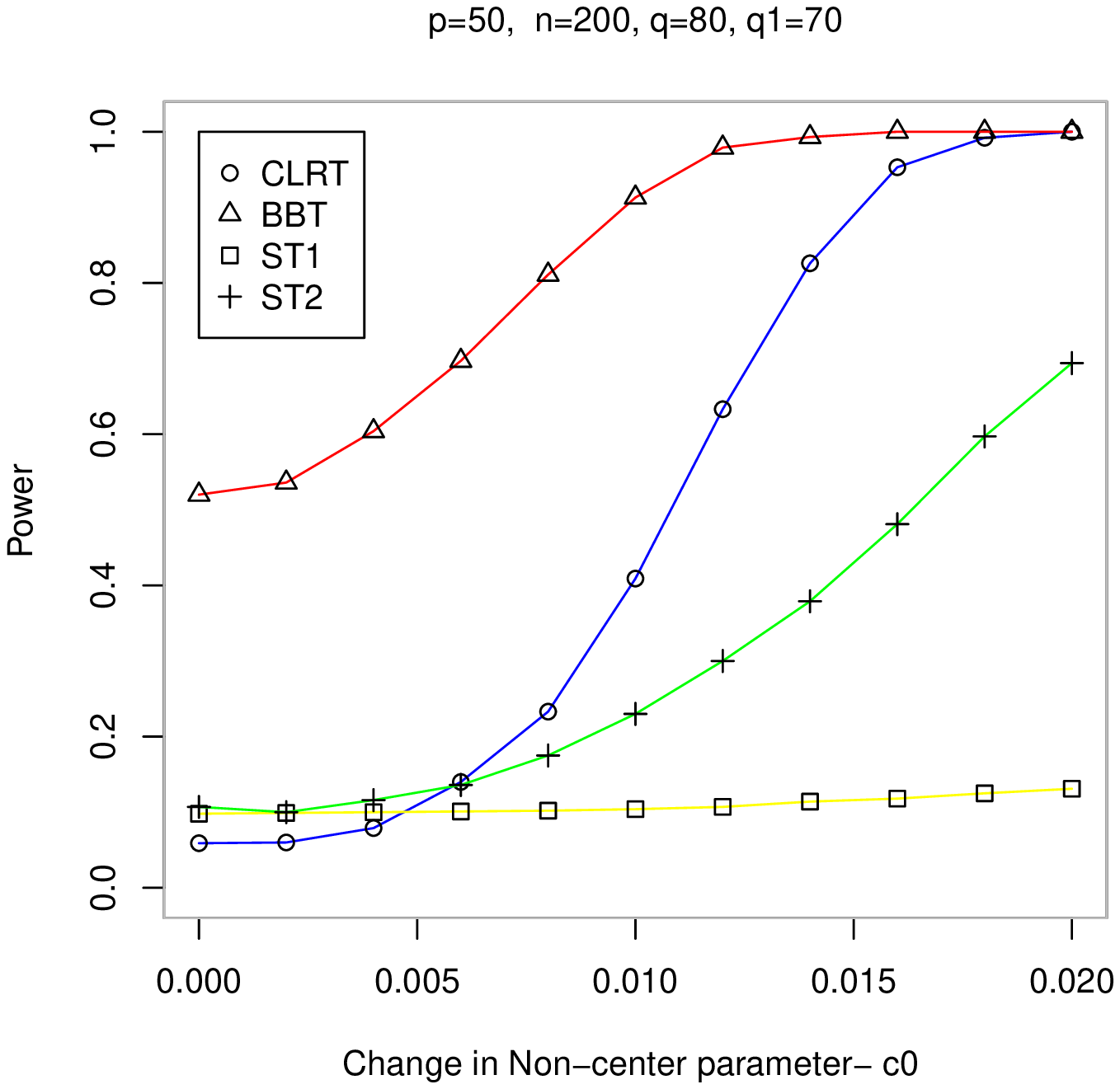}
    \caption{Sizes ($c_0=0$) and Powers ($c_0\ne 0$) of  the four methods, which are the corrected
      LRT (CLRT),  Bartlett-Box correction (BBC) and two least-squares type  tests ( ST1 and ST2 ),   based on 1,000 independent replications using
      Gaussian error variables from $\CN(0, C)$ with  the parameter $\rho=0.9$.\quad
      Top row:  ~~   $(p,n,q,q_1)=(10,100,50,30)$ and $(20,100,60,50)$.\quad
      Bottom row:~~$(p,n,q,q_1)=(30,200,80,60)$ and $(50,200,80,70)$.\quad
      \label{fig:34}}
  \end{center}
\end{figure}

\end{document}